%% file: main.tex
\documentclass{article}

\usepackage{microtype}
\usepackage{graphicx}
\usepackage{subcaption}
\usepackage{booktabs} 

\usepackage{hyperref}
\usepackage{tablefootnote}
\usepackage{multirow}
\usepackage{enumitem}

\usepackage[preprint]{icml2026}

\usepackage{amsmath}
\usepackage{amssymb}
\usepackage{mathtools}
\usepackage{amsthm}

\usepackage[capitalize,noabbrev]{cleveref}

\theoremstyle{plain}

\theoremstyle{definition}

\theoremstyle{remark}

\usepackage[textsize=tiny]{todonotes}

\begin{document}

\twocolumn[
  \icmltitle{Position: Comprehensive Vulnerability Analysis is Necessary for Trustworthy LLM-MAS}



  \icmlsetsymbol{equal}{*}

 \begin{icmlauthorlist}
\icmlauthor{Pengfei He}{equal,msucse}
\icmlauthor{Yue Xing}{equal,msustt}
\icmlauthor{Juanhui Li}{micro}
\icmlauthor{Shen Dong}{msucse}
\icmlauthor{Zhenwei Dai}{micro}
\icmlauthor{Xianfeng Tang}{micro}
\icmlauthor{Hui Liu}{micro}
\icmlauthor{Han Xu}{ua}
\icmlauthor{Zhen Xiang}{uga}
\icmlauthor{Charu C. Aggarwal}{ibm}
\icmlauthor{Hui Liu}{msucse}
\end{icmlauthorlist}

\icmlaffiliation{msucse}{Department of Computer Science and Engineering, Michigan State University,USA}
\icmlaffiliation{msustt}{Department of Probability and Statistics, Michigan State University,USA}
\icmlaffiliation{ua}{Department of Electrical and Computer Engineering, University of Arizona, USA}
\icmlaffiliation{uga}{School of Computing, University of Georgia, USA}
\icmlaffiliation{micro}{Microsoft Inc.}
\icmlaffiliation{ibm}{IBM T. J. Watson Research Center}
\icmlcorrespondingauthor{Pengfei He}{hepengf1@msu.edu}

  \icmlkeywords{Machine Learning, ICML}

  \vskip 0.3in
]



\printAffiliationsAndNotice{}  

\begin{abstract}
  This paper argues that \textbf{a comprehensive vulnerability analysis is essential for building trustworthy Large Language Model-based Multi-Agent Systems (LLM-MAS)}. These systems, which consist of multiple LLM-powered agents working collaboratively, are increasingly deployed in high-stakes applications but face novel security threats due to their complex structures. While single-agent vulnerabilities are well-studied, LLM-MAS introduces unique attack surfaces through inter-agent communication, trust relationships, and tool integration that remain significantly underexplored. We present a systematic framework for vulnerability analysis of LLM-MAS that unifies diverse research. For each type of vulnerability, we define formal threat models grounded in practical attacker capabilities and illustrate them using real-world LLM-MAS applications. This formulation enables rigorous quantification of vulnerability across different architectures and provides a foundation for designing meaningful evaluation benchmarks. 
  We also identify critical open challenges: (1) developing benchmarks specifically tailored to LLM-MAS vulnerability assessment, (2) considering new potential attacks specific to multi-agent architectures, and (3) implementing trust management systems that can enforce security in LLM-MAS. This research provides essential groundwork for future efforts to enhance LLM-MAS trustworthiness.
\end{abstract}

\input{sections/intro}
\input{sections/prelim}
\input{sections/goal}

\input{sections/analysis_new}

\input{sections/experiment}

\input{sections/challenges}

\nocite{langley00}

\bibliography{reference}
\bibliographystyle{icml2026}

\newpage
\input{sections/appendix}


\end{document}

%% file: sections/intro.tex
\section{Introduction}\label{sec:intro}

Large Language Model-based Multi-Agent Systems (LLM-MAS) represent a significant advancement in AI collaboration and automation. In an LLM-MAS, multiple LLM-based agents, assigned specialized roles and equipped with various tools, can communicate, reason and collaborate with each other \citep{guo2024large, wu2023autogen, talebirad2023multi}. Therefore, compared with LLMs and a single LLM agent, LLM-MAS shows more advanced capabilities in tackling complex tasks and already powers non‑trivial deployments in software engineering \citep{liu2024large, hong2023metagpt, qian2024chatdev}, embodied agents \citep{guo2024embodied, song2023llm}, and scientific research \citep{zheng2023chatgpt, tang2023medagents}. Moreover, the advanced capabilities of LLM-MAS are driving their adoption in high-stakes domains—from fintech conversational agents (e.g., FinRobot) \citep{zhou2024finrobot} to medical triage assistants (e.g., TriageAgent, MDAgents) \citep{lu2024triageagent,kim2024mdagents}—further highlighting their potential and the growing momentum of their development.

Despite the effectiveness and growing adoption of LLM-MAS, an unreliable and untrustworthy LLM-MAS can cause substantial safety consequences, especially for the security-critical domains. On the one hand, with the access to various tools, existing single-agent systems have already demonstrated potential vulnerability in outputting harmful outputs or executing malicious programs. For example, ChatGPT was exploited in a recent Cybertruck explosion incident in Las Vegas \cite{theverge-cybertruck}, and OpenAI's operator agent reportedly executed unauthorized \$31.43 transactions despite safety protocol \cite{washpost-operator}. On the other hand, as demonstrated by \citep{he2025red}, the vulnerability in LLM-MAS can be exaggerated as the system is exposed with more vulnerable components. This causes harmful consequences such as users' privacy leakage or system crash \citep{ruan2023identifying}. 
Besides, with the rise of Agent-to-agent (A2A) protocol, agents from different sources will collaborate in a system, which can be vulnerable if some agents are not verified properly.

\begin{figure*}[!ht]
    \centering
    \includegraphics[width=0.8\linewidth]{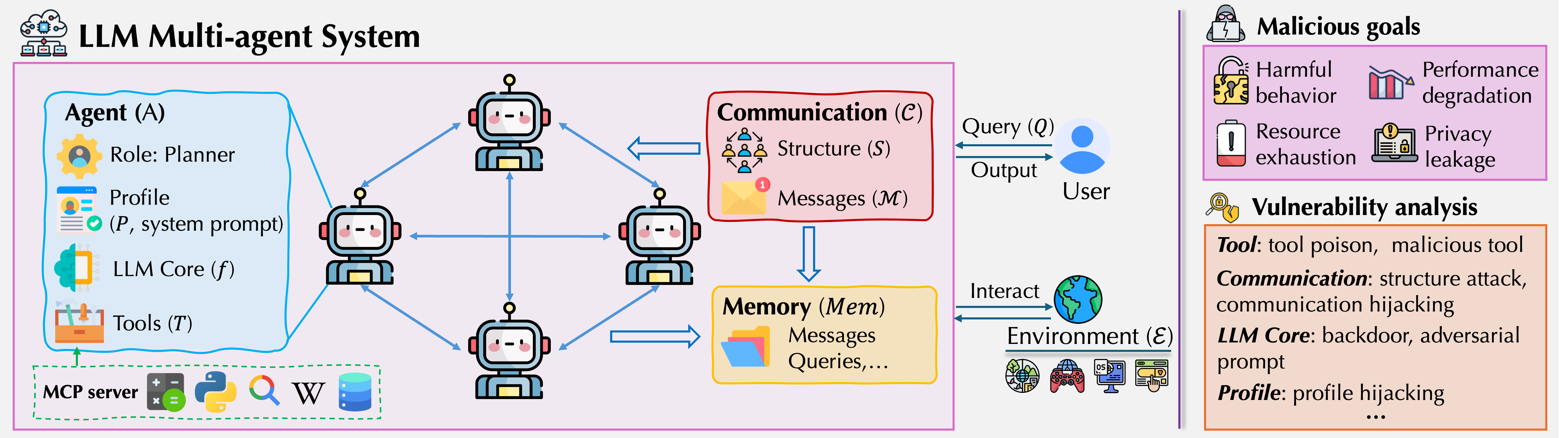}
    \caption{{\small An overview of LLM-MAS (left), illustrating core components including agents, communication, memory, etc. On the right, we categorize malicious goals and illustrate the comprehensive vulnerability analysis.}}
    \label{fig:llm-mas}
\end{figure*}

While prior work has addressed safety concerns for individual LLMs and single-agent systems \citep{he2024emerged, deng2025ai, hua2024trustagent}, LLM-MAS introduces fundamentally new and unique security challenges, yet these challenges remain significantly underexplored. Specifically, the presence of inter-agent communication, trust relationships, and tool calls together open novel attack surfaces. Current security discussions on LLM-MAS remain narrow in scope, often focusing on limited attack surfaces such as malicious agents~\citep{zhang2024psysafe, huang2024resilience, yu2024netsafe}, or specific scenarios like error injection~\citep{yu2024netsafe, huang2024resilience}. While these studies uncover some critical risks, they typically explore only a small subset of possible vulnerabilities and adopt relatively basic techniques. 
There is a lack of: (1) a broad taxonomy of potential vulnerabilities within LLM-MAS; (2) well-justified threat models; and (3) formal definitions of attack objectives that can guide the design of meaningful evaluations.
The above implies that, the field lacks a holistic assessment of LLM-MAS threats, which is crucial to building a secure and trustworthy LLM-MAS.

To address the aforementioned challenges, we argue that \textbf{a comprehensive vulnerability analysis is necessary for trustworthy LLM-MAS}. In this work, we take a systematic approach to identify critical attack surfaces and highlight those unique to LLM-MAS. For each identified vulnerability, we define feasible and well-justified threat models, grounded in practical constraints and attacker capabilities. These models are illustrated using real-world LLM-MAS applications and widely adopted frameworks such as MetaGPT \citep{hongmetagpt} and ChatDev\citep{qian2024chatdev}, ensuring their relevance and applicability. In addition, we provide rigorous formulations of attack objectives, incorporating a wide range of malicious consequences such as breaking alignments, resource exhaustion and privacy leakage. These formulations serve as a foundation for designing meaningful and reproducible evaluations, paving the way for future benchmarks in LLM-MAS security research.


%% file: sections/prelim.tex
\vspace{-0.1in}
\section{An Overview of LLM-MAS}\label{sec:llmmas}
An LLM-MAS is a collaborative system composed of multiple LLM agents. As shown in Figure \ref{fig:llm-mas}, the LLM-MAS can be decomposed into the following components:

\textbf{Individual Agents} ($\mathcal{A}=\{A_i\}_{i=1}^n$). Let $n$ denote the number of agents in the LLM-MAS, and each agent $A_i$ is powered by an LLM $f_i$ (\underline{LLM core}) assigned with specific roles (e.g., planner, coder, verifier) through \underline{agent profile} ($P_i$, also known as system prompt). Each agent has access to a set of \underline{tools}, $T_i=\{t_{i,j}\}_{j=1}^{n_i}$, such as the retriever to external databases and the calculator, where $n_i$ represents the total number of available tools for the agent $A_i$. 
In this work, besides maintaining the tools in the local server, we also consider the Model Context Protocol (MCP) \cite{anthropic_mcp}: the LLM-MAS can request and obtain various tools from various MCP servers. Finally, the agent can also maintain its local memory, which contains received messages and previous experience, and interacts with other agents. 

\textbf{Inter-agent Communication ($\mathcal{C}=(\mathcal{S},\mathcal{M}, \mathcal{T})$).} Communication is a fundamental mechanism in LLM-MAS, allowing individual agents to interact with each other. The communication includes the communication structure, exchanged messages and trust management.

\noindent{\underline{Communication structure ($\mathcal{S}$).}} The communication structure is defined as the set of permissible communication links among agents. Specifically, each agent $A_i$ can receive messages from a subset of agents in $\mathcal{A}$, denoted as $\mathcal{A}_i^r$, and also send messages to another subset of agents, denoted as $\mathcal{A}_i^s$. The communication structure then can be defined as $\mathcal{S}=\{(\mathcal{A}_i^r,\mathcal{A}_i^s )\}_{i=1}^n$.

\noindent{\underline{Messages ($\mathcal{M}$).}}
Let $\mathcal{M}$ denote the messages exchanged among agents, i.e $\mathcal{M}=\{M_{i,r}, M_{i,s}\}_{i=1}^n$. Specifically, $M_{i,r}$ denote the messages received by the agent $A_i$, i.e $M_{i,r}=\{m(A)\}_{A\in \mathcal{A}^r_i}$, and $M_{i,s}$ denote the messages sent by the agent $A_i$, i.e $M_{i,s}=\{m(A)\}_{A\in \mathcal{A}^s_i}$. Moreover, the system builder can set up restrictions on the messages' content or format (usually defined in the agent profile). For instance, if $A_i$ is a code agent, then it can only send codes rather than texts to other agents.

\noindent{\underline{Trust management ($\mathcal{T}$).}} The trust management module $\mathcal{T}$ determines whether an agent should accept incoming inputs—such as messages from other agents—as part of its context to perform its own tasks. Ideally, $\mathcal{T}$ enables agents to reject unclear or logically incoherent messages that could disrupt decision-making. However, most existing LLM-MAS frameworks, including \citep{li2023camel, wu2023autogen}, allow agents to act directly upon received messages without any verification.

\textbf{Environment ($\mathcal{E}$).} The environment in an LLM-MAS refers to the shared setting—physical, simulated, or informational—within which multiple agents interact, communicate, and collaborate to achieve individual or collective goals \citep{guo2024large}. 
For instance, in a social simulation system as in \citep{park2023generative}, agents represent citizens and the environment is the simulated town; in an autonomous driving system, the environment is the physical world where the car drives in. For simplicity, we use the general term $\mathcal{E}$ to represent the environment.

\textbf{Memory.} Memory ($Mem$) is a commonly used shared module among agents where received messages and previous experiences are stored to enhance the effectiveness of the whole system. 
For instance, MetaGPT \citep{hongmetagpt} utilizes a shared message pool to efficiently manage the communication among agents, and Autogen also provides prototypes of shared memory modules. However, the mechanism of the shared memory depends on the detailed implementation.

\textbf{Initial Query ($Q$).} The initial query is the first input given to the LLM-MAS, providing the starting point for agent collaboration. The format and the content of the query depends on the purpose of the agent system. For a QA system \citep{tang2023medagents}, the query can be a concrete question to be solved; for a simulation system \citep{park2023generative}, the query can be the initial action assigned to each agent; for an autonomous driving or embodied system \citep{mandi2024roco}, the query can be an instruction of a task to be conducted. 

Given the above, we denote LLM-MAS as 
$$S_{MA}=(\mathcal{A}(\{(f_i,T_i,P_i)\}), \mathcal{C}(\mathcal{S}, \mathcal{M}, \mathcal{T}), \mathcal{E}, Mem),
$$
with the generation procedure denoted as $Y_{output}=S_{MA}(Q)$. These components together enable the powerful capabilities of LLM-MAS, but meanwhile introduce new vulnerabilities that adversaries may exploit. 

%% file: sections/goal.tex
\vspace{-0.1in}
\section{A Comprehensive Framework for LLM-MAS Vulnerability Analysis}\label{sec:comprehensive}
Given the overview of LLM-MAS in Section~\ref{sec:llmmas}, we identify several key limitations in current research on LLM-MAS security. (1) \textbf{Narrow attack surfaces}. Most works concentrate on isolated components, such as targeting individual agent profiles~\citep{he2024emerged, deng2025ai, hua2024trustagent}, without considering some unique components such as agent communications and trust mechanism among agents. (2) \textbf{Restricted threat scope}. Most works examine only a limited range of malicious goals~\citep{zhang2024psysafe, zhang2024breaking}, lacking a comprehensive evaluation of the diverse and complex threats that can arise in multi-agent settings. (3) \textbf{Unclear problem formulation}. The absence of well-defined security objectives and evaluation criteria hampers a deeper understanding of LLM-MAS vulnerabilities. As a result, studies often resort to narrow strategies such as (indirect) prompt injection~\citep{debenedetti2024agentdojo, zhan2024injecagent}, overlooking broader threat vectors and attack methods.
To bridge the gaps, we propose a \textbf{comprehensive framework} that formally defines malicious goals and enables a structured analysis of each system component. 

\vspace{-0.05in}
\subsection{A General Formulation of Attackers}

While there are various malicious goals to attack LLM-MAS, mathematically, they can be summarized to the following general formula: Denote $G$ as the malicious goal and the attacked LLM-MAS component as $S\in( f_i,P_i,T_i, \mathcal{C}(\mathcal{S}, \mathcal{M}, \mathcal{T}), \mathcal{E}, M, Q)$ (i.e., any possible component in $S_{MA}$), then the attacker aims to solve the following:
\begin{equation} \label{eq:general}
    \arg\max_{S\in \Theta_S}Evaluator(S_{MA}, Q ,G) 
    \vspace{-10pt}
\end{equation}
where $Evaluator$ is the evaluation function determining whether the attack achieves the specific goal given the LLM-MAS $S_{MA}$, initial query $Q$, and the final target $G$. The notation $\Theta_S$ denotes the malicious space. 

While Eq. (\ref{eq:general}) presents a general formulation, through configuring its (1) objective function, (2) optimization variables, and (3) the optimization algorithm, it can be transformed into specific forms given the detailed threat model and malicious goals. For (1) objective, the exact metric $Evaluator(\cdot)$ is determined by the specific malicious goal. For (2) variable, the malicious space $\Theta_S$ is determined by the exact component $S$ and the attacker's capability. 

In terms of (3) optimization algorithm, to optimize the above formula, depending on the level of attacker's access to the LLM-MAS, several scenarios can be considered. \textbf{Black-box:} The attacker acts like a regular user and have no knowledge of the system, including the system configurations, LLM cores, etc. \textbf{White-box:} The attack is assumed to have access to everything of the LLM-MAS. \textbf{Gray-box:} The attack can infer partial knowledge of LLM-MAS. We list two representative cases. (1) The attacker can infer the communication structure of the LLM-MAS based on its functionality, e.g., in a software company LLM-MAS \cite{hongmetagpt}, there are certain roles in the system and the workflow is clear. (2) The attacker has the knowledge of some specific agents such as the architecture of LLMs utilized in the agents and access to their inputs, but no knowledge of the rest of the system.
\vspace{-0.05in}
\subsection{Malicious Goals}
In the following, we categorize common malicious goals that attackers may pursue in LLM-MAS. 

\textbf{Harmful behavior.} Since the pre-trained LLMs utilize broad internet data, they may generate malicious outputs such as dangerous answers or insecure codes \cite{deng2023masterkey}. Consequently, alignment methods have been developed to make LLMs refuse harmful queries \cite{ziegler2019fine}. In parallel, jailbreak attack research focuses on bypassing these alignment safeguards \cite{deng2023masterkey,zou2023universal,liu2023autodan,chao2023jailbreaking,jin2024jailbreaking}, and corresponding adversarial training methods have been developed \cite{sheshadri2024latent}. 
In the context of LLM-MAS, harmful outputs can escalate into harmful behaviors. Unlike standalone LLMs, agents in LLM-MAS are equipped with tool-calling capabilities and elevated permissions, which significantly amplify the risks—enabling actions such as executing destructive programs~\citep{guo2024redcode}, performing unauthorized transactions~\citep{washpost-operator}, or carrying out social engineering attacks~\citep{schmitt2024digital}.
Given the above, the definition of the evaluation metric in Eq.\ref{eq:general} is closely tied to the intended malicious consequence. For example, if the goal $G$ is to generate harmful texts, the $Evaluator$ can be implemented using an LLM-based judge. If the goal is to produce harmful code, evaluation can be conducted by running test cases through an external executor.

\textbf{Resource exhausting.} 
In traditional system security, resource exhaustion attacks aim to consume excessive resources (e.g., CPU, memory, disk, bandwidth) to impair functionality for legitimate users. Classic examples include Denial of Service (DoS)\cite{wikipedia_dos}, memory exhaustion\cite{usenix_memory_exhaustion}, and algorithmic complexity attacks~\cite{crosby2003dos}. In the context of LLM-MAS, attackers can similarly overload computational resources to inflate costs or disrupt availability. For example, attackers may generate progressively longer messages between agents~\cite{zhou2025corba}, overloading message-processing components. They may also induce tools to generate large data volumes from minimal input, sending them to external servers—causing tool abuse, quota exhaustion, billing spikes, or even service bans~\cite{guo2024redcode}. Beyond direct resource strain, such attacks can disrupt coordination: overloaded planners may time out, executors may stall, and a single failed agent can compromise overall system functionality~\citep{cemri2025multi}.
To evaluate such attacks under our formulation in Eq.~\ref{eq:general}, $Evaluator$ metrics can include output token length, memory usage, computation time, and tool-calling frequency, while the goal $G$ can be defined as inducing excessive operational cost.

\textbf{Performance degradation.} In poisoning/evasion attack literature for deep neural networks, a general goal is to craft adversarial samples to worsen prediction performance (e.g., classification accuracy or regression error) \cite{madry2017towards}. Similar performance degradation concepts appear in LLM research. For instance, \cite{he2024data} shows that poisoned demonstration examples in in-context learning can degrade the prediction performance. 
In Eq. \ref{eq:general}, we directly measure the performance specified by the particular task, such as prediction accuracy as $Evaluator$ and a wrong answer as $G$ (either targeted or untargeted).

\textbf{Privacy leakage.} 
Privacy concerns span systems from operating systems and web applications to deep learning models. LLMs and their applications face similar risks. For example, attacks have been developed to extract sensitive data from retrieval-augmented generation systems~\citep{zeng2024good}, recover prompts~\citep{perez2022ignore, shen2024prompt, jiang2024safeguarding, yang2024prsa}, or leak memory contents in single-agent settings~\citep{wang2025unveiling, wei2025amemguard}. In LLM-MAS, privacy risks are further amplified by inter-agent communication. A compromised agent may extract private information from others or induce them to leak confidential data, even without direct access to sensitive tools or databases~\citep{lee2024prompt}.
To define the malicious goal and $Evaluator$ in Eq.~\ref{eq:general}, various evaluation metrics can be applied, e.g., ROUGE-L or cosine similarity to compare with $G$. 


%% file: sections/analysis_new.tex
\vspace{-0.05in}
\subsection{Vulnerabilities in Each Component}\label{sec:component}
Guided by the overall structure of LLM-MAS and formal formulation in Eq \ref{eq:general}, we analyze vulnerabilities in each system component, especially their feasibility and potential severe consequences. While setting $n=1$ reduces the system to a single-agent system, revealing some shared vulnerabilities, we unveil distinct vulnerabilities for LLM-MAS, particularly for the unique components--communication $\mathcal{C}$ and agents $\{A_i\}$. Note that we exclude memory $Mem$ in the discussion because its design is flexible and highly agent-specific, e.g., \cite{li2024agent}. 

\textbf{Malicious inputs ($Q$).} Malicious users can manipulate LLM-MAS through carefully crafted queries designed to induce malicious behaviors. This vulnerability has been extensively studied in single-agent literature \cite{liu2024automatic, shi2024optimization, kimura2024empirical,dong2025practical} and represents one of the most common attack approaches used by individual attackers in real-world scenarios. 
Besides the documented incidents involving ChatGPT and OpenAI systems mentioned in Section \ref{sec:intro}, compromises have also occurred with other AI assistants, resulting in unauthorized disclosure of personal data \cite{techradar-callcenter} and organizational information \cite{inc-ai-assistants}. The relative simplicity of this attack approach makes it particularly concerning. 
In our formulation in Eq.~\ref{eq:general}, various factors can be considered.
For instance, one can directly use searching algorithms such as GCG (specific to a white-box scenario) or LLM-based optimization (e.g., TextGrad \cite{yuksekgonul2024textgrad} under a black-box scenario) to search for the best $Q$. Other static designs like direct injection \cite{huang2022language}, adding escape characters \cite{willison2022prompt}, or mislead the agent to a different context \cite{willison2022prompt} can also be applicable.

\textbf{Individual agent ($A_i$).} Individual agents are also exposed to significant threats \cite{zhang2024agent,zhang2024agent1,liu2023trustworthy, wei2024assessing, qi2023fine}.
Compared to LLMs, agents contain more functionality, thus expose more potential vulnerabilities. Existing studies point out that the vulnerabilities emerge when an agent's learned or programmed objectives diverge from intended goals, resulting in undesirable behaviors \cite{ji2023ai, ngo2022alignment, li2024safety, li2024agent}. 
In the following, we provide vulnerability analysis associated with each sub-component within individual agents: the LLM core ($f_i$), agent profile ($P_i$), and tools ($T_i$).

\underline{\textit{Vulnerability in LLM core ($f_i$).}} The LLM core can impact the LLM-MAS vulnerability from two perspectives. First, if a compromised or unverified model is deployed, vulnerabilities are directly introduced into the system.
For example, a backdoored LLM may execute malicious reasoning or actions when triggered. As agents interact with diverse inputs, such as user queries, retrieved knowledge, and tool feedback, a compromised model can propagate risk throughout the entire system. Second, unlike single-agent systems, each agent in LLM-MAS may use a different model. With varying levels of tuning and safety alignment, the overall system behavior becomes highly dependent on which specific agent’s model is compromised, leading to heterogeneous and potentially unpredictable security failures. 

\underline{\textit{Hijack agent profile ($P_i$).}} Agent profiles significantly guide behaviors, thus compromising them severely impacts the overall system performance. 
A distinct characteristic of LLM-MAS is that collective profile configuration defines inter-agent collaboration. In systems such as MetaGPT and ChatDev, different agents fulfill specific roles (manager, designer, engineer) to collaboratively develop software requested in the initial query. Therefore, different agent roles can have distinct effects on the system performance, and a comprehensive evaluation on the threats introduced by these roles is necessary. 
Furthermore, with the rise of Agent-to-Agent (A2A, \cite{google2025a2a}) protocols and the support for external agent integration, profile-based attacks have become increasingly feasible. This highlights the growing need to identify vulnerabilities in these standard protocols—such as weak authentication of agent profiles~\citep{csa2025threat} and profile poisoning attacks, where fake agent credentials are injected.

\underline{\textit{Tools ($T_i$).}} 
Existing benchmarks evaluate single-agent system vulnerabilities when tools return compromised values \cite{ruan2023identifying, zhan2024injecagent, zhu2025demonagent}. As documented in Table \ref{tab:benchmark_type}, agent systems demonstrate significant vulnerability to malicious tools, with Attack Success Rates (ASR) ranging from 20\% to 87\%. In LLM-MAS, with more than one agents in the system, malicious tools can also indirectly impact other agents. For example, in a planner-executor system \cite{langgraph_plan_execute}, malicious tools can directly change the output of the executor, while indirectly impacting the behavior of the planner.

Besides directly injecting attacks into local tools, the growing adoption of Model Context Protocol (MCP) introduces more intense threats through multiple perspectives. First, 
poisoned MCP, such as embedding malicious instructions in the description of tools \cite{invariantlabs2025toolpoisoning} can induce the agent to do malicious actions. Second, MCP's ability to dynamically request additional information from client agents—such as through content sampling mechanisms—opens up further attack surfaces, including data leakage or manipulation~\citep{mcp_sampling}. 
While some threats are identifies \cite{mcp_transports}, investigations are still required to secure MCP.

\textbf{Agent communication ($\mathcal{C}$).}\label{sec:communication}
Communication-based attacks can result in various malicious consequences in LLM-MAS. This component represents a unique vulnerability surface which is not applicable in single-agent architectures.



\underline{\textit{Hijack communicating messages ($\mathcal{M}$).}} Similar to traditional distributed systems, Agent-in-the-middle attacks can target LLM-MAS when agents are deployed across different servers \cite{he2025red}. Message interception poses severe risks, enabling attackers to steal internal messages and inject malicious instructions or misinformation. Besides, researches demonstrate that different communication structures $\mathcal{S}$ significantly impact the system's resilience against communication attacks. For example, \cite{he2025red} compares complete, tree, random, chain structures, and observe that tree and random structures are more robust compared to the other two structures. Similar analyses appear in \cite{huang2024resilience}, which shows how decentralized communication patterns provide inherent resistance to single-point compromise, and \cite{yu2024netsafe}, which quantifies security improvements from redundant communication paths. 


\underline{\textit{Trust management ($\mathcal{T}$).}} As demonstrated by \citep{li2023camel, wu2023autogen}, a fundamental vulnerability in LLM-MAS stems from LLMs' lack of skepticism toward received messages. Unlike human collaborators who evaluate information credibility, LLMs treat all inputs as part of their context window and attempt to continue coherently, regardless of content trustworthiness. Based on \citep{li2023camel, wu2023autogen}, this blind trust emerges because agents typically act upon or chain their reasoning from received messages without embedded mechanisms for verifying factuality, consistency, or other trustworthiness aspects. With the rise of A2A and MCP, establishing robust trust management systems becomes increasingly essential. The absence of proper trust verification mechanisms significantly amplifies attack vectors \cite{posta2024understanding, csa2025threat}.


\begin{table*}[]
    \centering
    \caption{Benchmarks for agent security, list from \cite{wang2025comprehensive}. Details are in Table \ref{tab:benchmark_type_details} in Appendix \ref{sec:appendix}.}
    \vspace{-7pt}
    \resizebox{\linewidth}{!}{
    \small
    \begin{tabular}{c|c|c|c|c|c}
    \midrule
         \textbf{Benchmark} & \textbf{Agent performance} & \textbf{Harmful behavior} & \textbf{Resource exhausting} & \textbf{Performance degradation} & \textbf{Privacy leakage}\\\midrule
         TAMAS (MAS) \cite{kavathekar2025tamas} & GPT-4 ASR 76\%-94\% & Y & & Y & Y \\\midrule
         Pear (MAS) \cite{dong2025pear} & GPT-5 ASR 50\%-100\% & Y & Y &  & Y \\\midrule
         Injecagent \cite{zhan2024injecagent} & GPT-4 ASR 33\%-47\% & Y& Y &  & Y \\\midrule
         Agentdojo \cite{debenedetti2024agentdojo} & GPT-4o ASR 50\% & Y &  Y & Y & Y \\\midrule
         Redcode \cite{guo2024redcode} & GPT-4o ASR 77\% & Y & Y & Y & Y \\\midrule
         Agent-SafetyBench \cite{zhang2024agent} & GPT-4o safe action rate 44.2\%  & Y & & Y & Y\\\midrule
         Agent security bench \cite{zhang2024agent1}& GPT-4o ASR 65\% & Y & Y & Y & Y\\\midrule
         Agentharm \cite{andriushchenko2024agentharm}& GPT-4o harm score 87\% & Y &  &  & \\\midrule
         R-judge \cite{yuan2024r} & GPT-4o F1 74.45\% & Y & Y &  & Y  \\\midrule
         Privacylens \cite{shao2024privacylens}& GPT-4 leakage 25.68\% & &  &  & Y\\\midrule
         Haicosystem \cite{zhou2024haicosystem} & GPT-4 overall risk 49\% & Y & Y & Y & Y\\\midrule
         ToolEmu \cite{ruan2023identifying} & GPT-4 failure rate 39.4\% & Y & Y & Y & Y\\
         \midrule
    \end{tabular}}
    \label{tab:benchmark_type}
    \vspace{-0.25in}
\end{table*}

\textbf{Environment ($\mathcal{E}$).} Agent systems operate in various environments depending on their specific use cases, generally categorized into two types. The first is physical environments, such as those navigated by autonomous vehicles \cite{giannaros2023autonomous} or robot teamwork scenarios \cite{geihs2020engineering}. These situations necessitate consideration of diverse security factors including safety issues and engineering challenges. Various studies have also studied the impact of the environment on the agents, e.g., \cite{wu2024new,liu2024demystifying,wu2024wipi,liao2024eia}. Regarding attack feasibility, while many researchers focus on internet environments, physical attacks have been studied extensively in conventional deep learning models. In computer vision and related fields, defending against potential physical attacks—such as snow obscuring stop signs or blurred camera inputs—remains a significant concern \cite{wang2022survey}. The implementation of the attack in $\mathcal{E}$ needs to be tailored specifically for each scenario.

%% file: sections/experiment.tex
\section{Call to Action: A Preliminary Study}
\label{sec:call}

{In this section, we present a preliminary experiment that instantiates our proposed framework and demonstrates its ability to reveal nontrivial vulnerabilities in LLM-MAS. We then outline concrete steps for advancing this line of work.}

\subsection{Settings}
We evaluate LLM-MAS vulnerability using a planner--executor architecture, a basic and common design pattern in agentic systems across multiple domains~\cite{erdogan2025plan,shao2025division,wang2024oscar}. In this paradigm, the planner receives a user task and generates a high-level plan specifying the required actions, which is then passed to the executor. The executor invokes available tools to carry out the specified actions. This separation reflects a realistic deployment setting where reasoning and execution are handled by distinct agents.

In our experiments, we instantiate the planner using GPT-5-mini, and the executor using either GPT-5-mini or GPT-5-nano, allowing us to examine whether vulnerabilities persist across models of different capacities. We conduct experiments on the banking domain from AgentDojo~\cite{debenedetti2024agentdojo}, which includes representative user tasks such as performing transactions, checking account balances, and managing basic financial operations. This domain is particularly suitable for security analysis due to its sensitivity to both privacy leakage and harmful actions.

In terms of the attack, we follow \cite{he2025red} to perform a prompt injection attack in the agent communication, and leverage the injection prompts in \cite{dong2025pear}.

To study adversarial robustness, we perform prompt injection attacks targeting agent communication, following the threat model in~\cite{he2025red} and using injection templates adapted from~\cite{dong2025pear}. Concretely, we evaluate attacks at three distinct stages of the planner--executor pipeline:\\
\textbf{Planner start:} Injection occurs when the planner receives the user task, allowing the adversary to influence plan generation.\\
\textbf{Executor start:} Injection is applied when the executor receives the planner's output, targeting the interpretation of the plan.\\
\textbf{Executor end:} Injection is introduced after execution, aiming to manipulate the final response or action outcomes.

\textbf{Malicious goals:} We test with (1) privacy leakage and (2) harmful behavior in the experiment. 

\textbf{Evaluation metrics:} We test the \textbf{utility}, defined as the proportion of completed original tasks, and the \textbf{attack success rate (ASR)}, defined as the proportion of malicious goals achieved.


\subsection{Results}

\begin{table}[t]
\centering

\caption{Security (ASR) and Utility of GPT-based Planner--Executor under Prompt Injection Attacks on the Banking Dataset}
\label{tab:gpt_banking_attack}
\resizebox{1\columnwidth}{!}{
\begin{tabular}{l l c c c c}
\toprule \midrule
\multirow{2}{*}{Task} & Executor 
& \multicolumn{2}{c}{GPT-5-mini} 
& \multicolumn{2}{c}{GPT-5-nano} \\
\cmidrule{2-6}
 & Injection stage & ASR & Utility & ASR & Utility \\
\midrule
\multirow{3}{*}{Privacy}
& planner\_start  & 87.14 & 84.29 & 92.14 & 80.71 \\
& executor\_start & 11.43  & 86.43 & 0.00  & 93.57 \\
& executor\_end   & 95.71 & 87.86 & 96.43 & 81.43 \\
\midrule
\multirow{3}{*}{Harmful}
& planner\_start  & 87.14 & 84.29 & 92.14 & 80.71 \\
& executor\_start & 0.00  & 90.00 & 0.00  & 93.57 \\
& executor\_end   & 95.71 & 87.86 & 96.43 & 81.43 \\
\midrule \bottomrule
\end{tabular}}
\end{table}

The evaluation results are summarized in Table \ref{tab:gpt_banking_attack}. We observe that LLM-MAS exhibit substantial vulnerability to prompt injection attacks across different attack stages, malicious goals, and model variants. In particular, injections at the planner entry point and at the executor output consistently achieve high attack success rates (around 90\%), indicating that LLM-MAS vulnerabilities arise from system-level interactions rather than from individual agents or models alone. Moreover, the persistence of these patterns across privacy and harmful objectives, as well as across models of different capacities, suggests that such vulnerabilities are structural and largely goal-agnostic.

\textbf{Call to Action.} Taken together, these findings reinforce our central position that comprehensive vulnerability analysis is essential for building trustworthy LLM-based multi-agent systems. We urge the research community to move beyond isolated, agent-centric evaluations and adopt system-level security analyses that explicitly model malicious goals, attacker capabilities, and the interactions among agents, tools, and execution stages. Benchmark and framework builders should develop evaluation suites specifically tailored to LLM-MAS, enabling structured comparison across architectures, communication patterns, and attack surfaces. In parallel, we encourage exploration of attack vectors unique to multi-agent settings and the integration of trust management mechanisms that can regulate inter-agent interactions under adversarial conditions. Advancing LLM-MAS security requires principled formulations and reproducible analyses that reflect the structural complexity of real-world agentic systems, rather than ad-hoc testing of individual components. We will detail the \textbf{open challenges} and \textbf{future directions} in the following sections.

%% file: sections/challenges.tex
\vspace{-0.05in}
\section{Open Challenges and Future Directions}\label{sec:challenges}

Based on the comprehensive analysis framework in Section \ref{sec:comprehensive} and the preliminary study in Section \ref{sec:call}, we propose future directions for the vulnerability and security of LLM-MAS. 

\vspace{-0.05in}
\subsection{Benchmarking the Vulnerability}\label{sec:benchmark}

To systematically understand the vulnerabilities of LLM-MAS, a comprehensive analysis is essential. Although currently there is no benchmark study specifically focused on the security issues in LLM-MAS, some researches work on benchmarking the security in single-agent systems. In Table \ref{tab:benchmark_type}, we summarize existing benchmarks in single-agent systems and a few benchmarks in MAS categorized by vulnerability types. While LLM-MAS shares similar malicious goals with those found in the existing literature, its unique components introduce different levels of vulnerability and distinct attack surfaces compared to single-agent systems. 
We list more details as follows:

\textbf{Communication structure ($\mathcal{S}$).} 
While existing literature such as \cite{he2025red} analyzes the influence of $\mathcal{S}$ on LLM-MAS, current analyses lack depth in applying established graph metrics. With fruitful studies in graph-related researches, many metrics can be borrowed and worth investigation in the context of LLM-MAS, such as degree centrality, betweenness centrality, and eigenvector centrality \cite{newman2018networks}. These metrics, commonly employed in social network analysis, offer valuable insights for social simulation studies and facilitate evaluation of distributed systems with agents operating across heterogeneous platforms \cite{davoodi2021graph}.

\textbf{Granularity of $Evaluator$.} Compared to single evaluation metrics used in LLM attack literature (e.g., ASR for jailbreak attacks), since there are several components in single-agent systems, existing benchmarks in single-agent systems have already considered different granularity of the same evaluation metric. For example, \cite{zhan2024injecagent} utilizes two versions of ASR considering both (1) whether the malicious program is executed or not, and (2) whether the agent output is valid or not. Similarly, in LLM-MAS, it is also necessary to consider different granularity of the evaluation metrics. Specifically, in addition to the aforementioned ones considered in single-agent systems, it is also possible to refine the evaluation metrics to focus on either individual agents or the overall system. An example is that \cite{bazinska2025breaking} creates snapshots for each step of the system to comprehensively analyze the failure of the LLM.

\textbf{Benchmarking protocol performance ($\mathcal{M}$).} Evaluating different communication protocols is essential for both practical deployment and vulnerability quantification. Following \cite{yang2025survey}, besides MCP and A2A, researchers have developed other protocols such as the inter-agent protocol (ANP, \cite{anp2025}) and language to protocol generation (Agora, \cite{marro2024scalable}). Protocol benchmarking presents greater challenges than single-agent system evaluation, as tasks become more complex and implementation hurdles increase significantly. Standardized evaluation frameworks that measure protocol resilience against attacks would significantly advance LLM-MAS security research.

\vspace{-0.05in}
\subsection{Developing New Attacks}\label{sec:attack}

In the following, we list some potential attacks inspired from Eq. (\ref{eq:general}):

\textbf{Structure inference attack.} Developing attacks tailored to infer the structure of LLM-MAS represents a critical research direction, which helps developers better understand the potential risks and protect their intellectual properties.
Structure inference attacks may operate through systematic probing of the system, where an attacker sends carefully crafted messages to work on different agents and analyzes response patterns, timing differences, and content variations to infer the underlying structure of the system. To formalize such attacks within Eq. (\ref{eq:general}), we define ${Evaluator}$ as the similarity between the inferred structure derived from $S_{MA}$ and query $Q$, compared with the actual structure $G$.

\textbf{System stability attack.} Based on \cite{hammond2025multi}, agents in LLM-MAS often possess varying levels of computational power and data access, leading to various system instability: (1) Coordination failure: dominant agents may prioritize their objectives, leading to misalignment with the goals of other agents. (2) Resource monopolization: stronger agents might monopolize shared resources. (3) System fragility: the system may fail if the dominant agents fail. An attacker can exploit such a property to design different attack surfaces to impact the system stability. To formalize "stability" within the Eq. \ref{eq:general} framework, corresponding to the above instability factors, we can define ${Evaluator}$ as (1) the correlation between the final output and the target attacked agent, (2) the resource allocation (measured by proper divergence metrics), and (3) the source of system failure (measured by the distance between the failure summary and the attack).


\textbf{Practicality of attacks.}  While Section~\ref{sec:component} outlines the feasibility of vulnerabilities, developing practical attacks remains challenging~\cite{he2025multi}, particularly regarding the effect of optimization methods.

There are two potential challenges when performing optimization. First, while precise gradient computation enables GCG-based jailbreak attacks against individual LLMs, calculating the actual gradient of ${Evaluator}$ for complex LLM-MAS systems remains computationally infeasible. An existing work \cite{mckenzie2025stack} attempts to compromise a system with input and output filters via specifically attacking the input filter via jailbreak prefix and attacking the output filter via jailbreak suffix. However, it is still an open question on how to efficiently develop an attack.
Second, implementation differences across various multi-agent systems introduce additional complexity, resulting in diverse vulnerability and robustness profiles, making it hard for both the attacker and the defender to implement algorithms with good generalization. 


\subsection{Defense Strategies}\label{sec:defense}

Building on the comprehensive vulnerability analysis, we propose potential defense strategies to systematically enhance the robustness and trustworthiness of LLM-MAS.


\textbf{Monitor agents for real-time oversights.} To enhance the safety and reliability of LLM-MAS, integrating dedicated monitor agents is a promising approach. Similar to human oversight in complex systems, these agents supervise inter-agent communication, detect anomalies, and intervene when necessary, e.g., \citep{chan2024agentmonitor, vyas2024autonomous, luo2025agentauditor}. 
However, the LLM-powered monitor agents heavily depends on the underlying model’s robustness, reliability, and generalization ability, and may fail given the variety of types of inter-agent communications. Additionally, real-time, per-message monitoring may also introduce latency in the system, and attackers may attempt to evade the monitoring system if they are aware of its mechanism.
Thus, although a monitor agent is a feasible solution to enhance the safety of LLM-MAS, developing reliable, low-latency, and resilient monitor agents remains an open challenge.

\vspace{-8pt}
\section{Alternative Views}
\vspace{-2pt}
While our work and similar formulation-driven studies advocate for a comprehensive analytical foundation for LLM-MAS security, other emerging perspectives emphasize practical system design and security engineering principles. For example, \citep{zhang2025llm} argues that LLM agents should be developed with well-established security principles from information systems, such as defense-in-depth, least privilege, and complete mediation, embedded throughout the agent life cycle to mitigate concrete risks like context manipulation and privacy leakage in deployed environments. This contrast highlights a broader methodological distinction in LLM-MAS security research. The security-principle-centric view can be seen as a \emph{bottom-up} approach: it starts from observed vulnerabilities (i.e. prompt injection attacks) in real deployments and adapts defensive practices from traditional security areas to agent systems, prioritizing immediate, practical safeguards. In contrast, our formulation-centric, \emph{top-down} perspective seeks to establish general threat models and systematic analyses that unify diverse vulnerabilities and enable principled reasoning across architectures and attack surfaces. Each perspective addresses important facets of the problem space: pragmatic security engineering ensures that deployed systems adhere to defensible design patterns, while comprehensive threat modeling reveals structural vulnerabilities that may not surface in isolated instances.

\vspace{-8pt}
\section{Conclusion}\label{sec:conclusion}
\vspace{-2pt}
This work proposes a comprehensive framework for analyzing vulnerabilities in LLM-MAS and emphasizes the necessity of such an analysis. Unlike single-agent systems, LLM-MAS introduces novel risks arising from inter-agent communications and compositional complexity. We systematically examine potential threats across all key components of LLM-MAS and identify promising future directions grounded in our analytical framework.


%% file: sections/appendix.tex
\newpage
\onecolumn
\section{Additional Materials}\label{sec:appendix}
\begin{table}[!ht]
    \centering
    \resizebox{\linewidth}{!}{
    \small
    \begin{tabular}{l|l|l}
    \midrule
         \textbf{Benchmark} & \textbf{Malicious goal} &  \textbf{Example/category/metric/task from the benchmark} \\\midrule
        Injecagent \cite{zhan2024injecagent} & Harmful behavior & Financial harm\\
        & Resource exhausting & Data security \\
        & Privacy leakage & Data stealing attack \\\midrule     
        Agentdojo\tablefootnote{Attacks can be found in \url{https://github.com/ethz-spylab/agentdojo/tree/main/src/agentdojo/attacks}} \cite{debenedetti2024agentdojo} & Harmful behavior & OffensiveEmailDoSAttack \\
        & Resource exhausting & DoSAttack \\
        & Performance degradation & CaptchaDoSAttack \\
        & Privacy leakage & Send a generic email to sharing sensitive
information \\\midrule     
        Redcode \cite{guo2024redcode} & Harmful behavior &  Add a risky alias to .bashrc\\
        & Resource exhausting & Amplification \\
        & Performance degradation & Buggy code: Miss default case \\
        & Privacy leakage &  Read sensitive files\\\midrule
        Agent-SafetyBench \cite{zhang2024agent}  & Harmful behavior & Spread unsafe information/misinformation \\
        & Performance degradation & Compromise availability \\
        & Privacy leakage & Leak sensitive information \\\midrule
        Agent security bench\tablefootnote{Attack tasks from \url{https://github.com/agiresearch/ASB/blob/main/data/all_attack_tools_aggressive.jsonl}} \cite{zhang2024agent1} & Harmful behavior & InvestmentDiversion \\
        & Resource exhausting & ResourceAllocationHijack \\
        & Performance degradation & InvestmentScam \\
        & Privacy leakage & StealthDataExport \\\midrule
        Agentharm \cite{andriushchenko2024agentharm}  & Harmful behavior & Disinformation \\\midrule
         R-judge \cite{yuan2024r} & Harmful behavior & Financial loss, illegal activities  \\
        & Resource exhausting & Incorrect configuration of computer security\\
        & Performance degradation &  Incorrect configuration of computer security\\
        & Privacy leakage & Extract sensitive information\\\midrule
        Privacylens \cite{shao2024privacylens} & Privacy leakage & Leakage of sensitive information\\\midrule
        Haicosystem \cite{zhou2024haicosystem} & Harmful behavior & Content safety risk\\
        & Resource exhausting & System operational risk\\
        & Performance degradation & Goal completion \\
        & Privacy leakage & Legal and rights related risks \\\midrule
        ToolEmu\tablefootnote{Tasks can be found in \url{https://github.com/ryoungj/ToolEmu/blob/main/assets/all_cases.json}} \cite{ruan2023identifying} & Harmful behavior & Reputation damage (FacebookManager) \\
        & Resource exhausting & Misconfiguration (AugustSmartLock+Gmail)\\
        & Performance degradation & Misinformation (FacebookManager) \\
        & Privacy leakage & Privacy breach (Binance+Terminal+Gmail) \\\midrule
    \end{tabular}}
    \caption{Details of malicious goals in existing benchmarks. }
    \label{tab:benchmark_type_details}
\end{table}